\documentclass[
twocolumn,
amsmath, amssymb, amsfonts,preprintnumbers,aps,prd,longbibliography,nofootinbib]{revtex4-2}

\usepackage{graphicx}
\pdfoutput=1

\usepackage{dcolumn}
\usepackage{bm}
 \usepackage{amstext}
 \usepackage{amssymb}
 \usepackage{amsmath}

 \usepackage{color}
 \usepackage{bbold}
 \usepackage{delimset} 
\usepackage[caption=false,justification=justified]{subfig}
\usepackage[colorlinks=true]{hyperref}
\usepackage{orcidlink}
\usepackage{float}
\usepackage[normalem]{ulem}
\usepackage{mathtools}
\usepackage[colorlinks=true]{hyperref}

\usepackage{tikz}
\usetikzlibrary{decorations.pathmorphing}
\usetikzlibrary{decorations.markings}
\usetikzlibrary{positioning, shapes, snakes}
\usetikzlibrary{arrows.meta}

\begin{document}

\author{Yong Zhang \!\orcidlink{0000-0002-3522-0885}}

\email{zhangyong1@nbu.edu.cn}

\affiliation{
Institute of Fundamental Physics and Quantum Technology\\ \& School of Physical Science and Technology, Ningbo University, Ningbo, Zhejiang 315211, China }

\title{Quantum fields from real-time ensemble dynamics}

\begin{abstract}
Relativistic quantum field theory (QFT) is commonly formulated in terms of operators,
asymptotic states, and covariant amplitudes, a perspective that tends to obscure
the real-time origin of field dynamics and correlations.
Here we formulate quantum fields in a real-time Schr\"odinger-picture framework,
in which fields evolve as probability ensembles on the space of field
configurations.
Within this formulation, the wavefunctional $\Psi[\phi,t]$ encodes a first-order,
causal ensemble dynamics on configuration space.
Interactions appear as couplings between configuration-space directions, while
propagators arise as derived correlation structures rather than as fundamental
postulates.
Entanglement, scattering amplitudes, and conformal field theory correlators emerge
as distinct projections of the same underlying ensemble evolution, corresponding
to equal-time, asymptotic, and symmetry-organized observables.
Standard operator, diagrammatic, and path-integral formulations are recovered as
computational representations of this single real-time dynamics.
This organization makes explicit the distinction between fundamental dynamical
structure and representational tools in QFT, and clarifies the
scope within which ensemble-averaged correlators account for quantum
fluctuations, while also delineating the level at which questions associated with
individual realizations and randomness would arise beyond the
correlator-based field-theoretic description.
\end{abstract}

\maketitle

\section*{Introduction}

Relativistic QFT provides the most successful framework
for describing fundamental interactions.
Its predictive power rests on a highly developed formalism based on operator
algebras, path integrals, and perturbative expansions, from which physical
content is extracted almost exclusively through correlation functions and
asymptotic observables~\cite{PeskinSchroeder,WeinbergQFT}.
In this standard organization, quantum fields are not primarily viewed as
objects evolving in real time, but rather as generators of correlators adapted
to scattering processes, symmetry constraints, and renormalization.

Despite its empirical success, this formulation tends to obscure the underlying
real-time dynamical structure of quantum fields.
The Schr\"odinger-picture description, in which a wavefunctional evolves on the
space of field configurations, is rarely emphasized in relativistic settings and
is often regarded as technically inconvenient.
As a result, fundamental notions such as causality, locality, and the dynamical
origin of correlations are typically inferred indirectly from properties of
propagators and commutators, rather than from an explicit description of
real-time evolution.

In this work we reformulate relativistic scalar field theory in a real-time
Schr\"odinger-picture framework, in which quantum states are described as
probability ensembles evolving on the configuration space of fields.
A single field configuration corresponds to a point in this space, while the
wavefunctional $\Psi[\phi,t]$ encodes an ensemble undergoing first-order time
evolution generated by a local Hamiltonian.
Within this formulation, interactions appear as couplings between directions on
configuration space, and propagators arise as derived correlation structures
rather than as fundamental postulates.

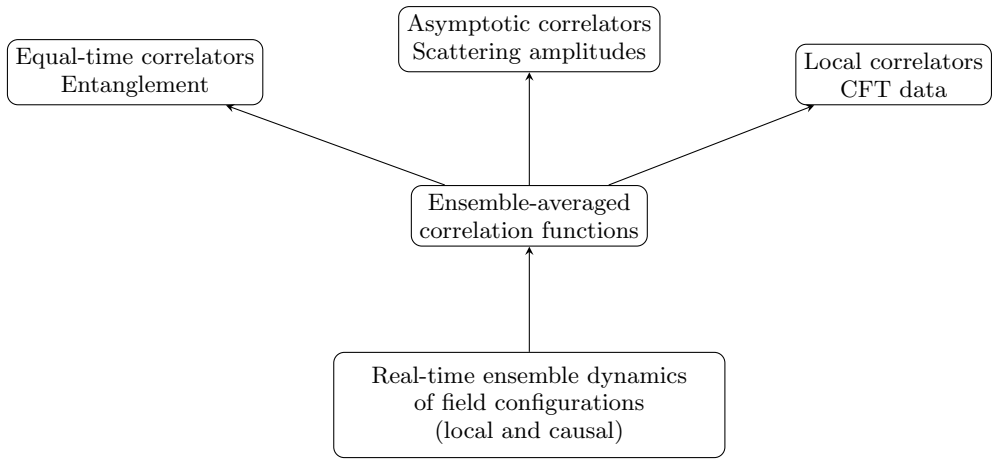
\begin{figure*}[t]
\centering
\begin{tikzpicture}[>=stealth, node distance=1.5cm]

\node[draw, rounded corners, align=center, minimum width=5.2cm, minimum height=1.4cm]
(core)
{
Real-time ensemble dynamics\\
of field configurations\\
{\small (local and causal)}
};

\node[draw, rounded corners, align=center, above=of core, yshift=-0.1cm]
(corr)
{
Ensemble-averaged\\
correlation functions
};

\node[draw, rounded corners, align=center, above left=of corr, xshift=-0.9cm]
(ent)
{
Equal-time correlators\\
Entanglement
};

\node[draw, rounded corners, align=center, above=of corr]
(amp)
{
Asymptotic correlators\\
Scattering amplitudes
};

\node[draw, rounded corners, align=center, above right=of corr, xshift=0.9cm]
(cft)
{
Local correlators\\
CFT data
};

\draw[->] (core) -- (corr);
\draw[->] (corr) -- (ent);
\draw[->] (corr) -- (amp);
\draw[->] (corr) -- (cft);

\end{tikzpicture}
\caption{
Organization of observables in the real-time ensemble formulation.
Local and causal ensemble dynamics on configuration space generates
ensemble-averaged correlation functions.
Different classes of physical observables arise from distinct projections,
limits, or symmetry constraints imposed on these correlators, including
equal-time entanglement measures, asymptotic scattering amplitudes, and
conformal field theory data.
}
\label{fig:ensemble_observables}
\end{figure*}

This ensemble-based organization provides a unified dynamical language for a
broad class of field-theoretic observables.
Equal-time correlators encode entanglement as an intrinsic property of the
ensemble.
Asymptotic limits of real-time evolution define scattering amplitudes.
Local correlators organized by spacetime symmetries give rise to conformal and
near-conformal data.
These observables represent different projections of the same underlying
ensemble dynamics, rather than independent ingredients of the theory, as
schematically illustrated in Fig.~\ref{fig:ensemble_observables}.

A particularly transparent realization of this viewpoint is obtained through a
Hamilton--Jacobi--Schr\"odinger (HJS) formulation, in which a linear complex
representation of classical Hamilton--Jacobi (HJ) ensemble dynamics is introduced~\cite{Zhang:2026gip}.
Within this representation, the wavefunctional obeys a first-order
Schr\"odinger-type equation, while linearity, superposition, canonical operators,
and the Born probability rule emerge as structural features of the
representation, rather than as independent quantization postulates.

Throughout this paper we emphasize a clear separation between fundamental
dynamical structure and representational tools.
At the fundamental level, the theory is defined by local, causal ensemble
evolution in real time.
Standard operator, diagrammatic, and path-integral formulations are recovered as
efficient computational representations adapted to particular classes of
observables, rather than as defining principles.

Finally, organizing QFT in this ensemble-based manner also
clarifies its structural relation to nonrelativistic quantum mechanics.
Both theories describe physical states at the level of ensembles, but differ in
the classes of observables they naturally emphasize.
While QFT is fundamentally organized around ensemble-averaged
correlation functions, the present formulation makes explicit the underlying
ensemble structure from which these correlators arise.
This perspective allows questions associated with individual realizations and
randomness—often emphasized in quantum-mechanical contexts—to be meaningfully
located relative to the correlator-based observables that define QFT, even though their detailed analysis lies beyond the scope of the present
work.

\section*{HJ ensemble dynamics}

We begin by formulating classical relativistic scalar field theory entirely
within the HJ framework.
The purpose of this section is not to introduce quantum structure, but to make
explicit the ensemble-theoretic and geometric organization of classical field
dynamics on configuration space.
This formulation will later provide the structural backbone for a linear
complex representation, but at this stage the discussion remains entirely
classical.

In the HJ description, a field configuration at fixed time is represented as a
point in an infinite-dimensional configuration space.
Classical dynamics corresponds to a deterministic flow on this space generated
by the HJ functional.
An ensemble of classical fields is then described by a density transported along
this flow.
No probabilistic or quantum interpretation is assumed at this level; the
ensemble description simply reorganizes classical field theory in a way that
treats families of solutions collectively rather than individually.

This ensemble-based formulation isolates the kinematical and dynamical
structures that will later admit a linear complex representation.
In particular, it cleanly separates the flow-defining functional from the
ensemble density on configuration space.
As we shall see, this distinction provides the natural starting point for the
subsequent organization of wavefunctionals and correlation functions on
configuration space, while remaining entirely classical in origin.

\section*{Linear completion and wavefunctionals}
\label{sec:LinearCompletion}

We now introduce a linear complex representation of classical scalar field
theory formulated at the level of HJ ensemble dynamics.
The purpose of this construction is structural rather than spectral.
Rather than analyzing particular states or excitations, our aim is to identify
the minimal quantum structures that arise when ensemble dynamics on
configuration space is represented in a linear and superposable form.

This linear completion provides a natural setting in which the familiar
wavefunctional Schr\"odinger equation, superposition principle, canonical
operators, and probabilistic interpretation of quantum fields can be recovered
as consequences of representation.
More generally, it establishes the structural basis for organizing quantum
states and correlation functions on configuration space, which will be explored
in subsequent sections.

\medskip

\paragraph*{Linear complex embedding of ensemble variables.}

As in the point-particle case, the nonlinearity of classical HJ ensemble dynamics
originates entirely from the quadratic momentum term.
For the scalar field, this nonlinearity takes the functional form
\begin{equation}
\int d^3x\,
\left(
\frac{\delta S[\phi,t]}{\delta\phi(\mathbf{x})}
\right)^2.
\end{equation}
Requiring a linear complex representation of the ensemble variables therefore
leads naturally to the wavefunctional embedding
\begin{equation}
\Psi[\phi,t]
=
R[\phi,t]\,
\exp\!\left(\frac{i}{\kappa}S[\phi,t]\right),
\label{eq:PsiDef_field}
\end{equation}
where $R^2=\rho$ is the ensemble density on configuration space and $\kappa$ is a
constant with dimensions of action.
At this stage, $\Psi$ is not postulated as a probability amplitude; it is simply
a complex encoding of the ensemble variables compatible with linear evolution.

\medskip

\paragraph*{Schr\"odinger-type evolution and quantum potential.}

Substituting the embedding~\eqref{eq:PsiDef_field} into the classical ensemble
equations yields the modified HJ equation
\begin{align}
\partial_t S[\phi,t]
+
\int d^3x
\Big[
\frac{1}{2}
\left(
\frac{\delta S}{\delta\phi(\mathbf{x})}
\right)^2
+
\frac{1}{2}(\nabla\phi)^2
&+
V(\phi)
\Big]
\nonumber
\\
&
+
Q[R]
=
0,
\label{eq:HJSfield}
\end{align}
with the functional quantum potential
\begin{equation}
Q[R]
=
-\frac{\kappa^2}{2}
\int d^3x\,
\frac{1}{R}
\frac{\delta^2 R}{\delta\phi(\mathbf{x})^2}.
\label{eq:Qfield}
\end{equation}
Together with the continuity equation for the ensemble density,
\begin{equation}
\partial_t \rho[\phi,t]
+
\int d^3x\,
\frac{\delta}{\delta\phi(\mathbf{x})}
\left(
\rho[\phi,t]\,
\frac{\delta S[\phi,t]}{\delta\phi(\mathbf{x})}
\right)
=
0,
\label{eq:HJfield_cont}
\end{equation}
this system is exactly equivalent to the linear Schr\"odinger-type equation
\begin{align}
i\kappa\,\partial_t\Psi[\phi,t]
=
\int d^3x
\Big[
-&\frac{\kappa^2}{2}
\frac{\delta^2}{\delta\phi(\mathbf{x})^2}
\nonumber
\\ &+
\frac{1}{2}(\nabla\phi)^2
+
V(\phi)
\Big]
\Psi[\phi,t].
\label{eq:SchField}
\end{align}
The Schr\"odinger equation, the quantum potential, and the statistical
interpretation of the ensemble thus arise at the same structural level from the
linear representation.

The above construction remains valid for a general complex deformation
parameter $\kappa$.
In the present work we restrict attention to real $\kappa$, corresponding to a
time-reversal invariant realization, in order to streamline the presentation and
focus on the core dynamical structure.

\medskip

\paragraph*{Superposition and probabilistic interpretation.}

Because the Schr\"odinger-type equation~\eqref{eq:SchField} is linear,
superposition of wavefunctionals follows automatically.
This linearity reflects the fact that $\Psi$ encodes ensemble information rather
than individual field configurations.

The ensemble density is naturally identified as
\begin{equation}
\rho[\phi,t]
=
|\Psi[\phi,t]|^2,
\label{eq:BornRule}
\end{equation}
reproducing the Born rule as the statistical interpretation of the
configuration-space ensemble.
The classical limit is transparent: as $|\kappa|\to0$, the quantum potential
vanishes and the ensemble dynamics reduces smoothly to classical HJ field
theory.

\medskip

\paragraph*{Operators as linear representations.}

Once the wavefunctional representation is fixed, familiar operator structures
can be recovered as linear representations of classical ensemble variables.
The HJ momentum field
\begin{equation}
\pi(\mathbf{x},t)
=
\frac{\delta S}{\delta\phi(\mathbf{x})}
\end{equation}
may be rewritten identically as
\begin{equation}
\pi(\mathbf{x},t)
=
-\,i\kappa\,
\frac{1}{\Psi}
\frac{\delta\Psi}{\delta\phi(\mathbf{x})}
\;-\;
\kappa\,
\frac{1}{R}
\frac{\delta R}{\delta\phi(\mathbf{x})}.
\end{equation}
Compatibility with superposition selects the linear part, leading to the
momentum operator
\begin{equation}
\hat\pi(\mathbf{x})
=
-\,i\kappa\,
\frac{\delta}{\delta\phi(\mathbf{x})}.
\label{eq:pihat}
\end{equation}
The remaining nonlinear contribution retains ensemble-dependent information and
remains part of the HJ description; it vanishes identically on eigenstates of
the momentum operator~\eqref{eq:pihat}.

With the field operator acting multiplicatively,
\begin{equation}
\hat\phi(\mathbf{x})\,\Psi[\phi]
=
\phi(\mathbf{x})\,\Psi[\phi],
\end{equation}
the equal-time canonical commutator follows directly,
\begin{equation}
\label{equal-time_canonical_commutator_of_pi_phi}
\bigl[\hat\phi(\mathbf{x}),\hat\pi(\mathbf{y})\bigr]
=
i\kappa\,\delta^{(3)}(\mathbf{x}-\mathbf{y}),
\end{equation}
without independent postulation.

\medskip

The same logic applies to the Hamiltonian.
Rewriting the classical Hamiltonian functional in terms of $\Psi$ yields a
decomposition into a linear operator and a nonlinear ensemble contribution.
The linear part defines the Hamiltonian operator
\begin{equation}
\hat H
=
\int d^3x
\left[
-\frac{\kappa^2}{2}
\frac{\delta^2}{\delta\phi(\mathbf{x})^2}
+
\frac{1}{2}(\nabla\phi)^2
+
V(\phi)
\right],
\label{eq:Hhat}
\end{equation}
which generates the Schr\"odinger-type evolution.
The nonlinear quantum potential \eqref{eq:Qfield} appears only in the HJ representation and does
not act on superpositions.

\medskip

\paragraph*{Expectation values and equal-time correlations.}

With the ensemble interpretation fixed, expectation values of functionals
$\mathcal O[\phi]$ are given by ensemble averages,
\begin{equation}
\langle \mathcal O \rangle
=
\int \mathcal D\phi\;
|\Psi[\phi,t]|^2\,
\mathcal O[\phi].
\end{equation}
At equal time, correlation functions admit a purely statistical interpretation.
In particular,
\begin{equation}
\label{equal_time_correlator}
\langle \hat\phi(\mathbf{x})\,\hat\phi(\mathbf{y}) \rangle
=
\int \mathcal D\phi\;
\rho[\phi,t]\,
\phi(\mathbf{x})\,\phi(\mathbf{y})
\end{equation}
characterizes the covariance structure of the configuration-space ensemble.
Dynamical propagation and causal structure enter through time-dependent
correlation functions, which we discuss next.

\section*{Real-time ensemble dynamics of quantum fields}

We now formulate relativistic QFT directly in terms of
real-time ensemble dynamics within the HJS framework.
The central object is the evolution of probability distributions on the space
of field configurations, governed by a first-order Schr\"odinger-type equation.
We begin with free scalar fields, for which the quadratic structure of the
Hamiltonian renders the ensemble dynamics exactly solvable.
This solvable setting provides a transparent arena in which vacuum structure,
field fluctuations, and propagators arise directly from real-time ensemble
evolution, without invoking particle excitations or second-order field
equations as fundamental postulates.
Interactions are then incorporated as couplings between configuration-space
directions, preserving linear real-time evolution while enriching the ensemble
structure.

\subsection*{Free-field ensemble dynamics}

We consider a real scalar field with free Hamiltonian
\begin{equation}
H_0[\phi,\pi]
=
\int d^3x\,
\left[
\frac{1}{2}\,\pi(\mathbf{x})^2
+
\frac{1}{2}(\nabla\phi)^2
+
\frac{1}{2}m^2\phi(\mathbf{x})^2
\right].
\end{equation}
The Hamiltonian consists entirely of quadratic terms and therefore defines a
linear flow on configuration space.
Within the HJS formulation, the corresponding ensemble dynamics is equivalently
encoded in a wavefunctional $\Psi[\phi,t]$ obeying the linear Schr\"odinger-type
equation
\begin{align}
&i\kappa\,\partial_t\Psi[\phi,t]
=
\hat H_0\,\Psi[\phi,t],
\nonumber
\\
&\hat H_0
=
\int d^3x
\left[
-\frac{\kappa^2}{2}
\frac{\delta^2}{\delta\phi(\mathbf{x})^2}
+
\frac{1}{2}(\nabla\phi)^2
+
\frac{1}{2}m^2\phi(\mathbf{x})^2
\right].
\end{align}

Because $\hat H_0$ is quadratic, the configuration-space directions decouple.
Equivalently, after Fourier decomposition, the ensemble factorizes into
independent directions labeled by momentum $\mathbf{k}$.
The free field therefore corresponds to a collection of decoupled quadratic
degrees of freedom on configuration space, each evolving independently under
linear real-time dynamics.

\subsection*{Vacuum ensemble and Gaussian structure}

The vacuum is defined as the stationary ensemble that minimizes the expectation
value of $\hat H_0$.
For each momentum mode, the problem reduces to that of a one-dimensional
harmonic oscillator, whose ground-state ensemble is uniquely Gaussian.
Assembling all modes, the vacuum wavefunctional takes the form
\begin{align}
\Psi_0[\phi]
=
\mathcal N
\exp\!\Big[
-\frac{1}{2\kappa}
\int\!\frac{d^3k}{(2\pi)^3}\,&
\omega_{\mathbf{k}}\,
\phi_{\mathbf{k}}\phi_{-\mathbf{k}}
\Big],
\nonumber
\\ 
&\omega_{\mathbf{k}}=\sqrt{\mathbf{k}^2+m^2}.
\end{align}

In the HJS decomposition $\Psi = R\,e^{iS/\kappa}$, the vacuum has constant phase,
\[
S[\phi]=\mathrm{const},
\qquad
\frac{\delta S}{\delta\phi(\mathbf{x})}=0,
\]
and therefore corresponds to a stationary ensemble with no HJ
flow.
The vacuum is not an empty field configuration, but a probability distribution
of minimal irreducible width on configuration space, fixed entirely by the
quadratic Hamiltonian.

The vacuum energy,
\begin{equation}
E_0
=
\int\!\frac{d^3k}{(2\pi)^3}\,
\frac{1}{2}\,\kappa\,\omega_{\mathbf{k}},
\end{equation}
arises directly from this finite ensemble width.
Its ultraviolet divergence reflects the accumulation of infinitely many
independent quadratic directions in configuration space, rather than unbounded
fluctuations in any single degree of freedom.

\subsection*{Propagators as dynamical correlations}

Having established the Gaussian vacuum ensemble, we now turn to time-dependent
correlation functions.
In the present formulation, propagators are not introduced as fundamental
objects.
They arise as dynamical correlations generated by real-time Schr\"odinger
evolution of the ensemble.

At equal time, correlation functions admit a purely statistical interpretation
and are given by ensemble covariances.
Time-dependent correlators probe how these fluctuations propagate under the
Hamiltonian flow on configuration space.

Operationally, the two-point function is defined by inserting field operators
into the ensemble and evolving the resulting correlations in time.
For $t>t'$, one has
\begin{equation}
G(\mathbf{x},t;\mathbf{y},t')
=
\int \mathcal D\phi\;
\Psi_0^{\ast}[\phi]\,
\hat\phi(\mathbf{x})\,
e^{-\,\frac{i}{\kappa}\hat H_0 (t-t')}\,
\hat\phi(\mathbf{y})\,
\Psi_0[\phi],
\end{equation}
with the ordering reversed for $t<t'$.
This may be written compactly as the time-ordered ensemble correlator
\begin{equation}
G(\mathbf{x},t;\mathbf{y},t')
=
\langle
T\,\hat\phi(\mathbf{x},t)\,\hat\phi(\mathbf{y},t')
\rangle,
\end{equation}
where time ordering is defined with respect to the physical Schr\"odinger time.

\subsection*{Emergent Green-function equation}

Because the free Hamiltonian is quadratic, Schr\"odinger evolution preserves the
Gaussian structure of the ensemble.
As a result, the two-point function satisfies a linear Green-function equation,
\begin{equation}
\left(
\partial_t^2
-
\nabla^2
+
m^2
\right)
G(\mathbf{x},t;\mathbf{y},t')
=
-i\kappa\,\delta(t-t')\,\delta^{(3)}(\mathbf{x}-\mathbf{y}).
\end{equation}
The resulting equation coincides with the standard Klein--Gordon Green-function
form, but here it arises as a derived property of ensemble dynamics rather than
as a fundamental dynamical postulate.

No independent second-order field equation is assumed.
The result follows solely from the following structural ingredients:
(i) linear Schr\"odinger evolution generated by the quadratic Hamiltonian
$\hat H_0$;
(ii) the definition of time-ordered ensemble correlation functions; and
(iii) equal-time commutativity of field configurations at distinct spatial
points,
\[
[\hat\phi(\mathbf{x},t),\hat\phi(\mathbf{y},t)] = 0 ,
\]
a direct consequence of the locality of configuration space, together with the
canonical equal-time commutation relation
\eqref{equal-time_canonical_commutator_of_pi_phi}.

\subsection*{Causality and locality}

In the Schr\"odinger representation adopted here, the field operator acts
multiplicatively on the wavefunctional,
\[
\hat\phi(\mathbf{x},t)\,\Psi[\phi,t]
=
\phi(\mathbf{x})\,\Psi[\phi,t].
\]
Equal-time commutativity of field operators at distinct spatial points therefore
follows directly from the configuration-space structure itself: field values
$\phi(\mathbf{x})$ and $\phi(\mathbf{y})$ correspond to independent coordinates
on configuration space for $\mathbf{x}\neq\mathbf{y}$.
This property is kinematical and does not rely on any additional assumptions
about dynamics.

Causal structure enters through the real-time evolution generated by the
Hamiltonian.
The Hamiltonian functional is local in space, coupling field values and their
spatial gradients only at the same spatial point.
As a result, the ensemble dynamics propagates information through neighboring
field configurations in a manner consistent with relativistic dispersion.
For free fields, this structure leads directly to the relation
$\omega_{\mathbf{k}}^2=\mathbf{k}^2+m^2$, which governs the spacetime support and
analytic properties of time-dependent correlations.

The same structural features persist in the interacting theory.
Interactions modify the ensemble dynamics by correlating configuration-space
directions that are independent in the free case, but the real-time evolution
remains generated by a local Hamiltonian.
The quantum potential $Q[R]$ arises only in the HJ representation of
the ensemble description and encodes the dependence of the dynamics on the local
structure of the ensemble density on configuration space.
It does not introduce additional dynamical variables or independent channels of
propagation beyond those already present in the linear Schr\"odinger evolution.

Statements about relativistic causality are therefore naturally formulated at
the level of time-ordered correlation functions constructed from real-time
evolution.
These correlation functions inherit their causal properties entirely from the
locality of the Hamiltonian dynamics.
Within this ensemble-based formulation, averaging over configurations does not
generate any new spacetime nonlocality; it organizes correlations already
produced by local and causal evolution.

\subsection*{Interacting fields as coupled ensemble evolution}

We now extend the real-time ensemble formulation to interacting scalar field
theories.
Interactions introduce no new kinematical ingredients.
Their sole effect is to modify the ensemble dynamics on configuration space by
correlating directions that are independent in the free theory.

We consider a scalar field with interaction potential
$V_{\mathrm{int}}(\phi)$, for example
$V_{\mathrm{int}}(\phi)=\lambda\phi^4/4!$.
The Hamiltonian takes the form
\begin{equation}
H
=
\int d^3x
\left[
\frac{1}{2}\,\pi^2
+
\frac{1}{2}(\nabla\phi)^2
+
\frac{1}{2}m^2\phi^2
+
V_{\mathrm{int}}(\phi)
\right].
\end{equation}

Unlike the free Hamiltonian, this expression no longer decomposes into
independent quadratic directions on configuration space.
The HJ flow generated by $H$ is therefore intrinsically coupled:
the evolution of the ensemble density depends on correlations among field values
at different spatial points.
From the ensemble perspective, this loss of factorization is the defining
structural effect of interactions.

Despite this increased complexity, the Schr\"odinger-type evolution equation
for the wavefunctional remains linear.
Interacting QFT is therefore not characterized by nonlinear time evolution,
but by the growing structural complexity of the ensemble distribution on
configuration space induced by interactions.

\subsection*{Vertices and perturbative structure}

Perturbation theory arises naturally by organizing the ensemble dynamics around
the Gaussian vacuum associated with the quadratic part of the Hamiltonian.
At leading order, the free propagator describes the linear response of the
ensemble to localized field insertions.
Higher-order contributions are generated by repeated action of the interaction
Hamiltonian, which correlates multiple configuration-space directions.

In this formulation, interaction vertices do not correspond to localized
particle collisions.
They encode how the interaction potential modifies the ensemble flow by coupling
several field degrees of freedom at a given time.
Their combinatorial structure reflects the polynomial form of
$V_{\mathrm{int}}(\phi)$, while their dynamical role is fixed entirely by the same
linear Schr\"odinger evolution that governs the free theory.

The privileged role of the free propagator thus follows directly from organizing
the ensemble expansion around a solvable Gaussian background.
It is not an additional assumption, but a structural consequence of perturbative
organization on configuration space.

\section*{Entanglement, asymptotic observables and CFT correlators}
Having formulated the real-time ensemble dynamics of quantum fields, we now turn
to physical observables.
In QFT, physical observables are typically organized at the
level of ensemble-averaged correlation functions generated by the underlying
dynamics, rather than being associated with individual field configurations.

Correlation functions characterize statistical fluctuations of the field
ensemble.
Their time dependence encodes how these fluctuations propagate, reorganize, and
become correlated under real-time evolution.
Different classes of observables correspond to different ways of organizing this
ensemble information.

Three such classes play a particularly prominent and largely independent role in
QFT.
The first consists of equal-time correlation functions, which characterize the
instantaneous structure of the ensemble and provide a natural framework for
discussing entanglement between different subsets of field degrees of freedom.
The second class involves asymptotic correlation functions evaluated between
in- and out-states, which encode scattering processes and give rise to the notion
of amplitudes.
A third class is provided by correlation functions in CFT,
which organize ensemble correlations according to exact spacetime symmetries
rather than asymptotic particle states or subsystem decompositions.

Although these observables are derived from the same underlying ensemble
dynamics, they probe complementary aspects of the theory.
Equal-time correlators retain explicit sensitivity to the local structure of the
ensemble at a given time.
Asymptotic observables arise after long-time evolution and projection onto
free-field dynamics.
CFT correlators, by contrast, extract symmetry-constrained information from the
ensemble without reference to scattering or particle interpretations.

In what follows, we discuss how entanglement, scattering amplitudes, and CFT
correlators each emerge naturally within the real-time ensemble formulation, and
how their standard field-theoretic descriptions are recovered as distinct but
compatible ways of organizing ensemble-averaged correlations.

\subsection*{Entanglement and equal-time correlators}

Within the real-time ensemble formulation, entanglement admits a direct and
intrinsic characterization as an equal-time property of the quantum state.
It is fully encoded in the correlation structure of the configuration-space
ensemble and is defined without reference to measurements, state reduction, or
asymptotic limits.

In QFT, nontrivial entanglement is already present
in the vacuum ensemble of a free field.
It arises from the impossibility of factorizing the configuration-space
distribution under a decomposition into subregions or subsets of degrees of
freedom.
Interactions modify but do not qualitatively alter this structure: they reshape
the ensemble correlations while preserving entanglement as an intrinsic feature
of the state.

\medskip

\paragraph*{Gaussian ensembles and subsystem structure.}

For a free scalar field, the vacuum is described by a Gaussian wavefunctional.
Its statistical content is therefore completely determined by equal-time
two-point correlators, which define the covariance structure of the ensemble
\eqref{equal_time_correlator}.
All higher-point correlations follow from these covariances.

Entanglement arises whenever the ensemble fails to factorize with respect to a
chosen subdivision of configuration space.
Restricting attention to a subregion and tracing out complementary degrees of
freedom produces a reduced ensemble that is generically mixed, even when the
global ensemble is pure.
This non-factorization is directly visible at the level of equal-time
correlation functions and requires no additional interpretational input.

\medskip

\paragraph*{Real-time evolution and causality.}

The Schr\"odinger-type real-time evolution of the wavefunctional induces a
deterministic evolution of ensemble correlations, and hence of the entanglement
structure.
This evolution is local and causal at the level of the underlying field
dynamics, even though correlations may extend over macroscopic distances.

From the ensemble viewpoint, entanglement reflects the global organization of
correlations within a causally evolving distribution on configuration space.
It does not correspond to any superluminal influence or dynamical nonlocality.
The real-time formulation makes this distinction manifest by keeping the
underlying dynamics local while allowing correlations to build up and reorganize
over time.

\subsection*{Asymptotic ensembles and scattering amplitudes}

Scattering amplitudes probe a complementary aspect of the ensemble dynamics:
they characterize correlations that persist after long-time real-time
evolution.
Whereas equal-time observables capture the instantaneous structure of the
ensemble, amplitudes encode how correlations generated during finite-time
interactions are reorganized and retained in the asymptotic regime.

Their definition relies on the existence of dynamical regimes in which the
ensemble evolution becomes effectively free.
In the present framework, this means that at early and late times the
wavefunctional approaches free-field evolution,
\begin{equation}
\Psi[\phi,t]
\;\xrightarrow[t\to\pm\infty]{}\;
\Psi_{\mathrm{free}}[\phi,t].
\end{equation}
This asymptotic decoupling follows from the dynamics itself and is not imposed as
an independent axiom.

\medskip

\paragraph*{Asymptotic ensembles.}

The asymptotic decoupling allows one to define incoming and outgoing ensembles by
factoring out the free evolution,
\begin{align}
\Psi_{\mathrm{in}}
=&
\lim_{t\to -\infty}
e^{i H_0 t/\kappa}\,\Psi[\phi,t],
\nonumber
\\ 
\Psi_{\mathrm{out}}
=&
\lim_{t\to +\infty}
e^{i H_0 t/\kappa}\,\Psi[\phi,t].
\end{align}
Scattering is thus identified with the map
\begin{equation}
\Psi_{\mathrm{in}}
\;\longrightarrow\;
\Psi_{\mathrm{out}},
\end{equation}
induced by finite-time evolution under the interacting Hamiltonian.
No reference to particles, operator states, or asymptotic Fock spaces is required
at this stage.

\medskip

\paragraph*{Scattering amplitudes as ensemble overlaps.}

Scattering amplitudes are defined as overlaps between the asymptotic ensembles,
\begin{equation}
\mathcal A
=
\langle \Psi_{\mathrm{out}} | \Psi_{\mathrm{in}} \rangle
=
\int \mathcal D\phi\;
\Psi_{\mathrm{out}}^{\ast}[\phi]\,
\Psi_{\mathrm{in}}[\phi].
\end{equation}
This expression is an inner product on configuration space.
Quantum effects enter solely through the finite ensemble width controlled by
$\kappa$.

From this viewpoint, amplitudes probe only a restricted slice of the full
ensemble dynamics.
They quantify how interaction-induced correlations reshape the asymptotic
configuration-space distribution, selected by the asymptotic projection.

\medskip

\paragraph*{Particle limit and LSZ reduction.}

The familiar particle interpretation emerges when the asymptotic ensembles
become sharply localized around a finite set of normal modes.
In this regime, the ensemble overlap reduces to a transition amplitude between
free excitations, with dominant contributions arising from connected
correlations.

Factoring out the universal free evolution already encoded in the asymptotic
dynamics leads naturally to amputated correlators.
Within the ensemble framework, amputation corresponds to removing the linear
propagation associated with the Gaussian background, leaving only the
correlations generated by interactions.

When expressed in terms of time-ordered correlation functions, this procedure
reproduces the Lehmann--Symanzik--Zimmermann (LSZ) reduction formula.
From the present perspective, LSZ is not a postulate about particles, but a
practical method for extracting asymptotic mode amplitudes from real-time
ensemble correlations.

\paragraph*{Covariance and integral representations}

Throughout the discussion above, the formulation has been expressed in the
Schr\"odinger picture, with time evolution defined relative to a chosen slicing.
Lorentz covariance is therefore not manifest at intermediate stages, reflecting
the use of real-time Hamiltonian evolution as the primary organizational
principle.

Physical observables of interest, however, are not defined at finite
intermediate times.
Scattering amplitudes are extracted only after asymptotic projection, once the
ensemble has evolved over the full spacetime domain.
At this stage, the explicit real-time evolution can be formally eliminated.

Operationally, this elimination proceeds by decomposing the finite-time evolution
operator into a sequence of infinitesimal time steps and composing the
corresponding short-time kernels.
Because the wavefunctional evolution is linear and first order in time, this
composition leads naturally to an integral representation over intermediate
field configurations, without requiring any assumptions beyond linear evolution
and the composition property of time evolution.

When written explicitly, the resulting expression takes the familiar
path-integral form,
\begin{equation}
\mathcal A
\;\sim\;
\int \mathcal D\phi\;
\exp\!\left[
\frac{i}{\kappa}
\int d^4x\,
\mathcal L(\phi,\partial_\mu\phi)
\right],
\label{eq:path_integral_like}
\end{equation}
where $\mathcal L$ is the Lagrangian density corresponding to the Hamiltonian that
governs the ensemble dynamics, and the functional integration is taken over field
configurations compatible with the chosen asymptotic boundary conditions.

The phase structure in Eq.~\eqref{eq:path_integral_like} is fixed by the linear
complex embedding $\Psi = R\,e^{iS/\kappa}$ underlying the ensemble formulation.
Once real-time evolution is composed and intermediate-time information is
eliminated, the appearance of the action functional divided by $\kappa$ follows
directly.

In this representation, dependence on any particular time slicing disappears.
Lorentz covariance is therefore recovered at the level of physical observables,
such as scattering amplitudes.

\subsection*{CFT correlators as symmetry-organized ensemble observables}

Correlation functions in CFT constitute an important class
of observables whose structure is fixed primarily by spacetime symmetries.
Within the ensemble formulation, such correlators admit a direct interpretation
as ensemble-averaged correlations evaluated in states whose real-time dynamics
is compatible with conformal invariance.

From this perspective, a CFT corresponds to an ensemble whose
Hamiltonian evolution preserves scale and conformal symmetry.
These symmetries impose strong constraints on the correlation structure of the
configuration-space distribution, restricting the allowed form of correlators
to a finite set of data, such as scaling dimensions and operator product
coefficients.

In the Schr\"odinger-picture formulation, equal-time correlators encode the
instantaneous covariance structure of the ensemble.
Imposing conformal invariance fixes how fluctuations are distributed and
correlated across configuration space in a scale-invariant state, independently
of any particle interpretation.
Higher-point correlators are then organized by symmetry rather than by dynamical
decomposition into elementary excitations.

Time-ordered CFT correlators arise from the real-time evolution of such
symmetry-constrained ensembles under a conformally invariant Hamiltonian.
Their characteristic spacetime dependence and analytic structure reflect the
combined effects of linear real-time propagation and conformal symmetry.
No reference to asymptotic states, scattering processes, or particle-like degrees
of freedom is required to define or interpret these observables.

From the ensemble viewpoint, conformal bootstrap conditions may be understood as
consistency requirements on the allowed correlation structures compatible with
both real-time evolution and conformal symmetry.
Crossing symmetry, unitarity bounds, and conformal block decompositions constrain
how ensemble correlations can be consistently organized, without modifying the
underlying dynamical framework.

In this way, CFT correlators are naturally interpreted as symmetry-organized
ensemble observables, providing a non-perturbative and representation-independent
characterization of QFT with conformal invariance.

\section*{Fundamental versus representational structures}

The formulation developed in this work highlights a clear separation between
fundamental dynamical structures and the representational frameworks commonly
used to express them.
While this distinction is often implicit in standard presentations of
QFT, it becomes explicit when the theory is
organized in terms of real-time ensemble dynamics on configuration space, where the fundamental object is the field ensemble itself, rather than any
particular choice of observables or representational framework.

\subsection*{First-order dynamics and determinism}

At the fundamental level, the dynamics is governed by first-order real-time
evolution.
Once the initial ensemble---specified by the HJ functional
$S[\phi]$ and the density $\rho[\phi]$---is fixed, the subsequent evolution is
fully determined by the modified HJ equation \eqref{eq:HJSfield} together with the
continuity equation.
In this sense, the ensemble dynamics is deterministic on configuration space,
even though individual outcomes inferred from ensemble observables remain probabilistic.

From this viewpoint, a wide class of physical questions can be traced back, in
principle, to properties of a single coupled dynamical system governing the
real-time evolution of configuration-space ensembles.
Quantum fluctuations, correlations, and observable processes do not require
independent postulates; they emerge as intrinsic properties of the evolving
ensemble itself.

Relativistic wave equations such as the Klein--Gordon and Dirac equations do not
appear as primary dynamical laws at the level of ensemble evolution in this
formulation.
Instead, they arise as familiar representational constructs at the level of
correlation functions, when the underlying ensemble dynamics is projected onto
time-ordered observables.
Such equations therefore provide differential descriptions of correlation
dynamics, rather than primary statements about the underlying real-time
evolution.

Locality and causality are already encoded in the classical Hamiltonian
structure and are preserved throughout the linear ensemble evolution.
Although Lorentz covariance is not manifest in intermediate Schr\"odinger-picture
expressions, it emerges unambiguously at the level of physical observables,
including correlation functions and scattering amplitudes.
Covariance is thus a property of observables, not a prerequisite for defining
the underlying dynamics.

Within this organization, the Schr\"odinger picture is not an auxiliary
representation but the fundamental dynamical language.
It provides a direct description of causal, real-time ensemble evolution on
configuration space.
Manifest Lorentz covariance is deferred rather than imposed, and appears
precisely when physical observables are extracted from the evolving ensemble.
(Related functional Schr\"odinger formulations of QFT have appeared in various contexts; see, e.g., standard textbook
discussions~\cite{Hatfield:2019sox}.)

\subsection*{Representational artifacts and calculational structures}

From this perspective, several structures commonly regarded as intrinsic
conceptual difficulties of QFT are revealed to be
representational rather than fundamental.
These include ultraviolet divergences, contour prescriptions such as the
Feynman $i\epsilon$, dimensional continuation, and the use of second-order
differential equations as defining dynamical laws.

Within the ensemble formulation, none of these structures is required at the
foundational level.
They arise when one adopts particular computational representations designed to
render specific calculations tractable.
Crucially, these representational structures are not tied to any single class of
observables.
They reflect general features of quantum field ensembles with infinitely many
degrees of freedom, and appear to different extents depending on which
correlation functions or limits are being probed.

A representative example is the divergence of vacuum zero-point energy. Within the present formulation, this divergence does not signal a physical
instability of the vacuum itself.
Rather, it reflects the breakdown of the free Gaussian description at
arbitrarily high energies, where interaction terms can no longer be neglected
and the quadratic approximation ceases to be valid.

More generally, ultraviolet divergences reflect the accumulation of infinitely
many independent directions on configuration space.
Regularization schemes and analytic continuation in spacetime dimension are
devices that control intermediate calculations, not statements about physical
spacetime itself.

A familiar analogy arises already in classical general relativity.
In the effective field theory treatment of gravitational-wave dynamics, loop
integrals, dimensional regularization, and an analytic continuation to
$4-2\epsilon$ spacetime dimensions are routinely employed as intermediate
computational tools (see, e.g., Ref.~\cite{Buonanno:2022pgc}).
Yet the same physical observables can be obtained within the traditional
post-Newtonian approach, formulated entirely in four-dimensional spacetime and
without invoking ultraviolet divergences or dimensional continuation.
 This coexistence illustrates that multiple representational frameworks can
faithfully encode the same underlying dynamics, while differing substantially
in their intermediate structures and calculational apparatus.

The same lesson applies in QFT. Different representations reorganize the same underlying ensemble dynamics in
ways adapted to different calculational tasks.
Operator methods provide an algebraic language for observables.
Diagrammatic perturbation theory offers a bookkeeping device for
interaction-induced correlations.
Path-integral representations emerge when real-time evolution is composed and
intermediate-time information is eliminated.
None of these frameworks introduces new physical principles.

\section*{Discussion}
\label{sec:Discussion}

In this work we have formulated QFT as real-time
ensemble dynamics on the space of field configurations.
Working in the Schr\"odinger picture, quantum fields are described as evolving
probability ensembles governed by a local Hamiltonian and first-order time
evolution.
This formulation makes the causal and local structure of the theory explicit
and provides a unified dynamical language for field fluctuations and
correlations.

Within this framework, we analyzed three complementary classes of observables.
Equal-time correlation functions characterize the instantaneous structure of
the ensemble and encode quantum entanglement as a property of
configuration-space correlations.
Asymptotic observables associated with scattering processes arise as overlaps
between free ensembles at early and late times, quantifying how
interaction-induced correlations accumulate during real-time evolution.
Local correlation functions at finite separations probe the short-distance and
symmetry-constrained structure of the ensemble and, in CFT, encode the operator content of the theory.

A key structural outcome of this organization is a clear separation between
fundamental dynamics and representational tools.
At the fundamental level, the theory is defined by deterministic ensemble
evolution on configuration space, governed by a local Hamiltonian.
Many structures commonly regarded as intrinsic to QFT —such as
second-order field equations, ultraviolet divergences, or specific calculational
prescriptions—arise only within particular representations adopted for
computational convenience.
From this perspective, the conceptual content of relativistic QFT resides in the structure and real-time evolution of ensembles on
configuration space, while the remaining challenges are primarily dynamical and
computational.

\subsection*{Scope, limitations, and open structural questions}

The present work has focused on relativistic scalar field theory as a minimal
and conceptually clean setting in which the real-time ensemble formulation can
be developed explicitly.
While extensions to fermionic fields and gauge theories are structurally
natural, a fully explicit treatment of constraints, gauge redundancy, and
fermionic functional measures lies beyond the scope of this paper.

Several open questions concern the extension of the ensemble framework to
regimes where additional structural ingredients become essential.
Although the ensemble viewpoint clarifies the origin of ultraviolet divergences
and the organization of perturbative expansions, it does not by itself resolve
the problem of vacuum energy in gravitational contexts.
The coupling of configuration-space ensembles to dynamical spacetime geometry
therefore remains an open challenge at the interface of QFT and quantum gravity.

Closely related issues arise in the context of spontaneous symmetry breaking.
Within an ensemble-based formulation, symmetry breaking is naturally associated
with the global structure and support of the configuration-space distribution,
rather than with the selection of a particular operator vacuum.
A detailed dynamical account of how degenerate minima, vacuum manifolds, and
symmetry-breaking patterns emerge from ensemble evolution warrants further
investigation.

Beyond these structural questions, the present work has emphasized three
distinguished classes of observables—equal-time correlators associated with
entanglement, asymptotic observables defining scattering amplitudes, and local
correlators characteristic of conformal field theories.
These observables correspond to particularly natural projections of the same
underlying real-time ensemble dynamics.

More general correlation functions also exist.
In particular, finite-time, finite-separation correlators in interacting,
nonconformal field theories probe genuinely dynamical regimes that are neither
captured by equal-time entanglement measures nor reducible to asymptotic
scattering data.
Such observables play a central role in nonequilibrium dynamics,
thermalization, and strongly coupled systems, but currently lack a comparably
unified structural organization.
The present framework does not provide explicit solutions for these correlators,
but offers a natural setting in which they can be defined and interpreted
directly as ensemble-averaged quantities generated by real-time evolution.

A particularly important instance of this broader class arises in QFT at finite temperature.
Because the present framework treats classical and quantum field dynamics
uniformly in terms of ensembles, thermal field theory admits a transparent
interpretation as mixed ensemble dynamics on configuration space.
This perspective opens the possibility of formulating finite-temperature and
nonequilibrium QFT directly in real time, without relying on imaginary-time
constructions, and of treating thermodynamic, entropic, and
information-theoretic quantities on the same footing as dynamical observables.

\subsection*{Ensembles, correlators, and randomness in QFT}

Relativistic QFT has historically been formulated as a theory
of correlation functions.
Its predictive content is extracted from time-ordered correlators, asymptotic
matrix elements, and symmetry-constrained correlation data.
Within this organization, quantum fields are not interpreted as providing
descriptions of individual events, but rather as generators of ensemble-averaged
statistical structures.
Single realizations of a field configuration do not appear as primitive objects
in the standard formulation.

This emphasis is reflected in both the conceptual and computational architecture
of QFT.
Operator algebras, path integrals, and perturbative expansions are all designed
to produce correlators adapted to scattering processes, symmetry constraints,
and renormalization.
Questions concerning individual detection events or single-shot outcomes are
typically addressed only after restricting to effective, few-degree-of-freedom
descriptions, or are delegated to nonrelativistic quantum mechanics.
At the level of fundamental field theory, the correlator is the primary carrier
of physical information.

The ensemble-based formulation developed in this work sharpens this structure.
Within the real-time Schr\"odinger-picture framework, quantum states are
described as probability ensembles evolving on the space of field
configurations.
Correlation functions arise as ensemble-averaged quantities constructed from
this underlying dynamics.
In this sense, correlators encode statistical properties of the ensemble as a
whole, rather than properties of individual realizations.
The ensemble itself, however, contains additional structure beyond its averaged
correlation data.

This distinction clarifies the structural location of randomness in a
field-theoretic setting.
If QFT is to accommodate notions of randomness associated with
individual events, such randomness cannot be encoded at the level of
ensemble-averaged correlators.
Correlators characterize fluctuations of the ensemble, but they do not specify
which particular configuration within the ensemble is realized in a given
instance.
Any notion of single-event randomness must therefore correspond to the selection
of a specific realization from the ensemble, rather than to a modification of
the correlator-level dynamics.

The present work does not attempt to construct or analyze the dynamics governing
individual realizations.
Its aim is instead structural.
By making explicit the distinction between ensemble evolution and ensemble
averaging, the real-time formulation delineates the level at which QFT makes definite physical predictions and the level at which questions of
randomness would have to be posed.
Establishing this separation is a necessary step for any further discussion of
randomness in relativistic field theory, even though it is not by itself
sufficient to resolve such questions.

\section*{Acknowledgements}
The work of Y.Z. is supported by the National Natural Science Foundation of China under Grant No. 12405086.

\bibliographystyle{physics}
\bibliography{Refs}

\end{document}